\documentclass[12pt]{article}

\usepackage{amsmath}

\usepackage{a4wide,cite}

\usepackage{epsfig,psfrag}

\numberwithin{equation}{section}
\def\openone{\leavevmode\hbox{\small1\kern-3.8pt\normalsize1}}%

\DeclareMathOperator{\ch}{ch}

\DeclareMathOperator{\Tr}{Tr}

\DeclareMathOperator{\length}{length}

\begin{document}
\baselineskip=21pt
\parskip=7pt


\hfill April 28, 1999

\vspace{24pt}

\begin{center}
  {\large\textbf{Supersymmetric   Polychronakos  Spin Chain:\\
      Motif, Distribution Function,  and Character.}}

  \vspace{24pt}

  Kazuhiro \textsc{Hikami}
  \footnote[1]{E-mail:
    \texttt{hikami@phys.s.u-tokyo.ac.jp}
    }
  and
  B. \textsc{Basu-Mallick}
  \footnote[2]{E-mail:
    \texttt{biru@monet.phys.s.u-tokyo.ac.jp}
    }

  \vspace{8pt}
  \textsl{Department of Physics, Graduate School of Science,\\
    University of Tokyo,\\
    Hongo 7--3--1, Bunkyo, Tokyo 113--0033,
    Japan
    }

  \vspace{8pt}
(Received: May 3, 1999)

\end{center}

\vspace{18pt}

\begin{center}
  \underline{ABSTRACT}
\end{center}
Degeneracy patterns and hyper-multiplet structure in the spectrum
of the  su($m|n$) supersymmetric Polychronakos spin chain
are studied by use of the ``motif''.
Using the recursion relation of the supersymmetric Rogers--Szeg{\"o}
polynomials which are closely related to the  partition function of
the  $N$ spin chain, we give  the
representation for motif in terms of the supersymmetric skew Young diagrams.
We  also  study the distribution function for quasi-particles.
The character formulae for $N\to \infty$ are briefly discussed.

\vfill

\newpage

\section{Introduction}

The integrable spin chain with an  inverse square interaction has
received extensive studies in these years.
One of the famous model is the Haldane--Shastry (HS)
spin  chain~\cite{Hal88,Sha88}.
The    su($m$)  HS  spin chain is integrable, and
contrary to the ordinary Heisenberg spin chain it has the
$\mathcal{Y}(s\ell_m)$ Yangian  symmetry
even in a finite chain~\cite{HHTBP92}.
While the quasi-particle (spinon)
for the Haldane--Shastry has a square  dispersion relation,
there exists another   integrable spin chain whose quasi-particle
has a linear dispersion relation.
This model, which we call  the Polychronakos   spin chain,
was  originally  introduced in
Ref.~\citen{Poly93a},
and thanks to a linear dispersion relation we can compute the
partition function~\cite{Frah93,Poly94a}.
What is remarkable  is that,
similar to the case of HS model,
the  Polychronakos  spin chain also possesses the
$\mathcal{Y}(s\ell_m)$ Yangian symmetry~\cite{Hikam94d}.

{}From the viewpoint of the conformal field theory, it was
realized~\cite{BPS94,Schou94}
that the Yangian  symmetry can be embedded into the level-1 WZNW
theory, and that the first Yangian invariant operator is the Virasoro
generator $L_0$.
Indeed, this  supports the fact that
the energy spectrum  for the 
Polychronakos  spin chain is equally spaced.
As the Polychronakos  spin chain  has the exact  Yangian symmetry
in the case of finite $N$ spins,
the character  for the level-1 WZNW theory can be given  in a large $N$ 
limit of the partition function of the Polychronakos spin
chain~\cite{Hikam94d}.
In this sense, the partition function of spin chain with finite number 
of lattice sites
is viewed as the  \emph{restricted} character formula.
In a computation of such  partition function,
the Yangian invariant bases which are  called the
``motifs''~\cite{HHTBP92,Hal94} play an important role.
These  motifs  span  the Fock space of the Yangian invariant system, and
the degeneracy for each motif can be given from  the
representation theory of the Yangian algebra.
The representation for the motif was given for su(2) case in the
original paper~\cite{HHTBP92}, and
for su($m$) case it was established by one of the
author~\cite{Hikam94d,Hikam94e},
where  the main observation is that the partition function for 
finite-site Polychronakos spin chain can be  identified with the
Rogers--Szeg{\"o} (RS)
polynomial~\cite{Andre76}  at some special values.

In this paper  we consider the su($m|n$) supersymmetric extension of the
Polychronakos  spin chain (SP model), whose Hamiltonian is written as
\begin{equation}
  \label{Hamilton_P}
  \mathcal{H}^{(m|n)}
  =
  \sum_{1 \leq i< j \leq N}
  \frac{
    1 - P_{i j}
    }{
    ( z_i - z_j )^2
    } .
\end{equation}
Here  $z_i$ (for $i=1,2,\dots,N$)
are  zeros of the $N$-th Hermite polynomial.
The supersymmetric  spin operator $P_{i j}$ is written as~\cite{Hal94}
\begin{equation}
  P_{i j}
  =
  \sum_{\alpha , \beta = 1}^{n+m}
  c_{i, \alpha}^\dagger \, c_{j, \beta}^\dagger  \,
  c_{i , \beta} \, c_{j, \alpha} ,
\end{equation}
Here  the  creation--annihilation operators, $c_{i,\alpha}^\dagger$ and 
$c_{i, \alpha}$, are
\begin{equation*}
  c_{i, \alpha}
  :
  \begin{cases}
    \text{bosonic} & \text{for $\alpha= 1, \dots, m$},
    \\[2mm]
    \text{fermionic}
    & \text{for $\alpha= m+1, \dots, m+n$ ,}
  \end{cases}
\end{equation*}
and we have a constraint
\begin{equation*}
  \sum_{\alpha=1}^{m+n} c_{i, \alpha}^\dagger \, c_{i, \alpha} = 1 .
\end{equation*}
Note that the spin operator $P_{ij}$ is a permutation operator,
satisfying
\begin{align*}
  P_{ij} \, P_{jk}
  &= P_{jk} \, P_{ki} = P_{ki} \, P_{ij}  ,
  &
  P_{ij}^{~2}
  & = 1 .
\end{align*}

As was shown in Ref.~\citen{Poly94a}, the su($m$)  Polychronakos spin
chain is
a  static   limit of the spin  Calogero  model confined in the harmonic 
potential.                
We can then compute the partition function of the su($m$)
Polychronakos spin model  by factoring out the
dynamical degree of freedom from the spin Calogero model.
By using a similar approach for the case of SP
model~\eqref{Hamilton_P},
one can obtain
the corresponding  partition function
$\mathcal{Z}_N^{(m|n)}(q)
= \Tr q^{\displaystyle \mathcal{H}^{(m|n)}}$
as~\cite{BiruUjinWada99a}
\begin{equation}
  \label{partition}
  \mathcal{Z}_N^{(m|n)}(q)
  =
  \sum_{
    \substack{
      \sum_{i=1}^m a_i + \sum_{j=1}^n b_j =N
      \\
      a_i \geq 0 , \quad b_j \geq 0
      }
    }
  \frac{
    (q ; q)_N
    }{
    \displaystyle
    \prod_{i=1}^m ( q ; q)_{a_i} \cdot
    \prod_{j=1}^n ( q  ; q)_{b_j}
    }
  \cdot
  q^{\displaystyle
    \frac{1}{2}
    \sum_{j=1}^n  b_j \, ( b_j - 1)
    }  .
\end{equation}
See \S~\ref{sec:pre} for definitions.
We note that the partition function~\eqref{partition} has the duality
for $m, n \neq 0$,
\begin{equation*}
  \mathcal{Z}_N^{(m|n)}(q)
  =
  q^{\displaystyle
    \frac{N \, (N-1)}{2}
    } \,
  \mathcal{Z}_N^{(n|m)}(q^{-1}) .
\end{equation*}
We remark  that in  the pure fermion case, \emph{i.e.}, when $m=0$,
we need to multiply prefactor in~\eqref{partition}
for taking into account  the non-zero  ground state energy of the
related spin Calogero model~\cite{Poly94a,BiruUjinWada99a}, and
in the following we always suppose $m \neq 0$.
A main purpose of the present Article
is to introduce   motifs    as eigenstates
of~\eqref{Hamilton_P}, and to give   representation  for the motifs.
We can naturally define the
supersymmetric RS  polynomial from the partition
function~\eqref{partition}, and
based on the recursion relation for those  polynomials
we can compute the degeneracy of motifs and the distribution function
for the quasi-particles.

This Article is organized as follows.
In \S~\ref{sec:pre} we  explain notations used in this paper.
We review  some important properties for the Schur polynomials
following Refs.~\citen{Macdo95,WFult97Book}.
In \S~\ref{sec:RSpoly} we introduce a supersymmetric analogue of the
Rogers--Szeg{\"o} polynomial.
This polynomial
would reproduce   the partition
function~\eqref{partition} at some  special values, and we study the
recursion relation for
these polynomials.
In \S~\ref{sec:representation}
we introduce  motif  as eigenstates of the
SP model, and
we give the representation for
motif using the supersymmetric skew Schur polynomials.
In \S~\ref{sec:distribute} we calculate the distribution function for
quasi-particles and
the central charge  by
use of the recursion relation for the restricted partition function.
We also discuss a relationship with the   character formula  in
\S~\ref{sec:WZW}.
The last section is devoted to discussions and concluding remarks.

\section{Preliminaries}
\label{sec:pre}
\subsection{Notation}
We denote  a $q$-polynomial as
\begin{equation}
  (t ; q)_N
  =
  \prod_{i=1}^N
  ( 1 - t \, q^{i-1} )  ,
\end{equation}
for $N>0$, and we set $(t;q)_0 = 1$.
For this polynomial, we have an identity~\cite{Andre76},
\begin{equation}
  \label{help_Eq}
  \frac{1}{
    (t ; q)_\infty
    }
  =
  \sum_{N=0}^\infty
  \frac{
    t^N
    }{
    (q ; q)_N
    } .
\end{equation}

A partition $\lambda$
(see Refs.~\citen{Macdo95,WFult97Book} for detail)
is given by a sequence of weakly decreasing
positive integers, and we often write it as
$\lambda = [ \lambda_1 , \lambda_2 ,\dots, \lambda_m]$.
The conjugate diagram $\Tilde{\lambda}$ is given by flipping a diagram 
$\lambda$ over its main diagonal (from upper left to lower right).

A skew diagram $\lambda/\mu$
is  obtained by removing a smaller Young diagram
$\mu=[\mu_1, \mu_2, \dots]$ from a 
larger one $\lambda$
that contains it.
Hereafter, we often use  the  skew Young diagrams
$\langle m_1 , m_2 , \dots, m_r \rangle$,
which denote the border strip of $r$-columns such that the length of
the $i$-th column is $m_i$;
\begin{equation}
  \label{border_strip}
  \langle m_1, m_2, \dots , m_r \rangle = 
  \begin{array}{*{4}{|p{\arraycolsep}}|ccc}
    \cline{4-4}
    \multicolumn{3}{c|}{} & & \uparrow & & \\
    \cline{4-4}
    \multicolumn{3}{c|}{} & & & & \\
    \cline{4-4}
    \multicolumn{3}{c|}{}   & \vdots & m_1 & & \\
    \cline{3-4}
    \multicolumn{2}{c|}{} & & &  \downarrow & \uparrow & \\
    \cline{3-4}
    \multicolumn{2}{c|}{} &
    \vdots & \multicolumn{1}{|c}{} & & m_2 & \\
    \cline{2-3}
    \multicolumn{1}{c|}{}  & & &\multicolumn{1}{|c}{}  & & \downarrow
    & \uparrow    \\
    \cline{2-3}
    \multicolumn{1}{c|}{} & \vdots & \multicolumn{2}{|c}{} & & & m_3 \\
    \cline{1-2} &  &  \multicolumn{2}{|c}{} & & & \downarrow\\
    \cline{1-2}
    \vdots & \multicolumn{6}{|c}{} \\
  \end{array}
\end{equation}

\subsection{Schur Polynomial}
For each partition  $\lambda$,
we define the  su($m$) Schur polynomial
$s_\lambda(x) = s_\lambda(x_1, \dots, x_m)$
by the Jacobi--Trudi
formula~\cite{Macdo95,WFult97Book};
\begin{equation}
  s_\lambda(x)
  =
  \frac{
    \det 
    \left(
      (x_j)^{\lambda_i + m - i}
    \right)_{1 \leq i , j \leq m}
    }{
    \det
    \left(
      (x_j)^{m - i}
    \right)_{1 \leq i , j \leq m }
    }  .
\end{equation}
Note that $s_\lambda(x)=0$ when $\length(\lambda) > m$.
For our brevity, we define
the elementary and  complete  symmetric polynomials $e_N(x)$ and
$h_N(x)$  respectively   as
\begin{align}
  e_N(x)
  & = s_{[1^N]}(x) ,
&
  h_N(x)
  & = s_{[N]}(x) .
  \label{element_complete}
\end{align}
The polynomial $h_N(x)$ is the sum of all distinct monomials of degree 
$N$ in the variables $x$, while $e_N(x)$ is the sum of all monomials
$x_{i_1} \dots x_{i_\ell}$ for all strictly increasing sequences
$1 \leq i_1 < \dots < i_\ell \leq m$.
We have the generating functions for  polynomials $e_N(x)$ and
$h_N(x)$ as
\begin{gather}
  \prod_{i=1}^m  ( 1 +  t \, x_i)
  =
  \sum_{N=0}^m  e_N(x) \, t^N ,
  \label{generate_element}
  \\[2mm]
  \prod_{i=1}^m
  \frac{1}{
    1 - t \, x_i
    }
  =
  \sum_{N=0}^\infty h_N(x) \, t^N .
  \label{generate_complete}
\end{gather}
We note that the Schur polynomials $s_\lambda(x)$ can be also given
as~\cite{WFult97Book}
\begin{align}
  \label{SchurDet}
  s_\lambda (x)
  & =
  \det
  \left(
    h_{\lambda_i + j - i} (x)
  \right)_{1 \leq i , j \leq \length(\lambda)}
  \nonumber \\
  & =
  \det
  \left(
    e_{\Tilde{\lambda}_i + j - i} ( x)
  \right)_{1 \leq i , j \leq \length(\Tilde{\lambda})}  .
\end{align}

The above definitions for the Schur polynomials can be generalized
straightforwardly to the
skew Young diagram $\lambda/\mu$.
We define the skew Schur polynomials by~\cite{Macdo95,WFult97Book}
\begin{align}
  s_{\lambda/\mu} (x)
  & =
  \det
  \left(
    h_{\lambda_i - \mu_j + j - i}(x)
  \right)_{1 \leq i , j \leq \length(\lambda)}
  \nonumber \\
  & =
  \det
  \left(
    e_{\Tilde{\lambda}_i - \Tilde{\mu}_j + j - i} (x)
  \right)_{1 \leq i , j \leq \length(\Tilde{\lambda})}  .
  \label{skewSchur}
\end{align}
where $\Tilde{\lambda}$ and $\Tilde{\mu}$ denote the conjugate
partitions.

\subsection{Supersymmetric Schur Polynomial}
The su($m|n$) supersymmetric analogue  of the Schur polynomials can be
defined as
follows.
As an extension of~\eqref{SchurDet},
we define the supersymmetric Schur polynomials
(sometimes called as the hook Schur polynomial, or bisymmetric
polynomial)
$S_\lambda(x,y) = S_{\lambda}(x_1, \dots, x_m , y_1 , \dots, y_n)$
as
(see, e.g., Ref.~\citen{WFult97Book})
\begin{equation}
  S_\lambda ( x , y)
  =
  \det
  \left(
    c_{\lambda_i + j - i}
  \right)_{
    1 \leq i, j \leq \length(\lambda)
    }  ,
\end{equation}
where $c_N = c_N(x,y)$ is given by
\begin{equation}
  \label{defineCN}
  \frac{\displaystyle
    \prod_{j=1}^n ( 1 + t \, y_j )
    }{
    \displaystyle
    \prod_{i=1}^m ( 1 - t \, x_i)
    }
  =
  \sum_{N=0}^\infty
  c_N \, t^N   .
\end{equation}
The coefficients $c_N$ correspond to  a  supersymmetric analogue of the
complete symmetric polynomial $h_N(x)$~\eqref{generate_complete}.
One easily sees a duality;
\begin{equation*}
  S_\lambda(x_1, \dots, x_m, y_1, \dots , y_n)
  =
  S_{\Tilde{\lambda}}(y_1, \dots, y_n , x_1 , \dots, x_m)  ,
\end{equation*}
where $\Tilde{\lambda}$ is the conjugate partition.

We can also  define the supersymmetric analogue
of the elementary  symmetric polynomials.
For our later convention, we set the supersymmetric elementary
polynomials as
\begin{equation}
  \label{susyElement}
  E_N(x,y) 
  =
  S_{[1^N]}(x,y)   .
\end{equation}
Based on identities~\eqref{generate_element},~\eqref{generate_complete}, 
and~\eqref{defineCN},
the   supersymmetric elementary function is
decomposed~\cite{BalanIBars82a,BarMorRue83a} as
\begin{equation*}
  E_N(x,y)
  =
  \sum_{k=0}^N
  e_{k}(x) \cdot h_{N-k}(y) .
\end{equation*}
By using this relation it is easy to see that
$E_N(x,y) \neq 0$ for arbitrary $N$.
We note that the generating function for the polynomials $E_N(x,y)$ is 
given by
\begin{equation}
  \label{generateSuperE}
  \frac{\displaystyle
    \prod_{i=1}^m ( 1 - t \, x_i )
    }{
    \displaystyle
    \prod_{j=1}^n ( 1 + t \, y_j)
    }
  =
  \sum_{N=0}^\infty (-)^N \, E_N(x,y) \, t^N .
\end{equation}

\section{Recursion Relation for the Partition Function}
\label{sec:RSpoly}

As the Rogers--Szeg{\"o} polynomial reproduces the partition function
for the su($m$) Polychronakos model at some special values~\cite{Hikam94d},
it is natural to study a supersymmetric analogue of the RS polynomials 
which is  related with the partition function~\eqref{partition}.
In this section
we  give the recursion relations for such supersymmetric RS
polynomials in terms
of the Young diagrams.

\subsection{Supersymmetric Rogers--Szeg{\"o} Polynomials}

We define 
polynomials
$H_N^{(m|n)}(x,y) = H_N^{(m|n)}(x_1, \dots, x_m, y_1, \dots, y_n ; q)$ 
as
\begin{equation}
  \label{polyH}
  H_N^{(m|n)}(x,y)
  =
  \sum_{
    \substack{
      \sum_{i=1}^m  a_i + \sum_{j=1}^n b_j = N
      \\
      a_i \geq 0 , \quad b_j \geq 0
      }
    }
  \frac{
    (q;q)_N
    }{
    \displaystyle
    \prod_{i=1}^m (q ; q)_{a_i} \cdot
    \prod_{j=1}^n (q^{-1} ; q^{-1} )_{b_j}
    }
  \cdot
  x_1^{a_1} \dots x_m^{a_m} \cdot
  \left(
    - \frac{y_1}{q}
  \right)^{b_1} \dots
  \left(
    - \frac{y_n}{q}
  \right)^{b_n} ,
\end{equation}
which we call the su($m|n$) supersymmetric Rogers--Szeg{\"o} (SRS)
polynomial.
It is easy to see  that the polynomial $H_N^{(m|n)}(x,y)$ gives the
partition function~\eqref{partition}
for the SP model~\eqref{Hamilton_P},
\begin{equation}
  \mathcal{Z}_N^{(m|n)}(q)
  =
  H_N^{(m|n)}(x = 1 , y=1)  .
\end{equation}

Next we try to derive  the recursion  relation for the SRS
polynomials~\eqref{polyH}.
For this  purpose,  it is useful to give  a generating function of the
polynomial~\eqref{polyH}.
So we introduce   a function
$G^{(m|n)}(t) = G^{(m|n)}(t ; x_1, \dots, x_m , y_1, \dots, y_n ;q )$
as            
\begin{equation}
  \label{generate_Function}
  G^{(m|n)}(t)
  =
  \frac{1}
  {
    \displaystyle
    \prod_{i=1}^m ( t \, x_i ; q)_\infty
    \cdot
    \prod_{j=1}^n
    ( - t \, y_j \, q^{-1} ; q^{-1} )_\infty
    }    .
\end{equation}
We see that,
by use of an identity~\eqref{help_Eq},
$G^{(m|n)}(t)$
is a generating function for the
SRS  polynomial~\eqref{polyH};
\begin{equation}
  \label{generateH}
  G^{(m|n)}(t)
  =
  \sum_{N=0}^\infty
  \frac{
    H_N^{(m|n)}(x,y)
    }{
    (q ; q)_N
    } \cdot
  t^N  .
\end{equation}
By definition~\eqref{generate_Function},
the function $G^{(m|n)}(t)$ satisfies
$q$-difference equation,
\begin{equation}
  \label{differenceG}
  \left(
    \prod_{j=1}^n
    (1 + t \, y_j)
  \right) \cdot
  G^{(m|n)} (q \, t)
  =
  \left(
    \prod_{i=1}^m
    (1 - t \, x_i)
  \right) \cdot
  G^{(m|n)} (t) .
\end{equation}

Substituting~\eqref{generateH} to above equation  and using the
elementary symmetric polynomials~\eqref{generate_element},
we see that the  SRS polynomial satisfies the following
recursion relation for any ($\max(n,m) +1$)-consecutive polynomials;
\begin{equation}
  \label{FirstRecur}
  H_N^{(m|n)}(x,y)
  =
  \sum_{k=1}^{\max(n,m)}
  (-)^{k-1}
  \frac{
    (q ;q )_{N-1}
    }{
    (q ; q)_{N-k}
    } \cdot
  \left(
    e_k(x) 
    - (-)^k \, q^{N-k} \, e_k(y)
  \right) \cdot
  H_{N-k}^{(m|n)}(x,y)  ,
\end{equation}
where
we have set
$H_k^{(m|n)}(x,y)=0$ for $k<0$.
However,  by using the expression~\eqref{generateSuperE} which contains
supersymmetric elementary  polynomials, we can rewrite the
$q$-difference equation~\eqref{differenceG} as
\begin{equation*}
  G^{(m|n)} (q \, t)
  =
  \left(
    \sum_{N=0}^\infty (-)^N \, E_N(x,y) \, t^N
  \right) \cdot
  G^{(m|n)} (t) .
\end{equation*}
Substituting~\eqref{generateH} to above equation,
we obtain  another type of recursion relation for the SRS polynomials
as 
\begin{equation}
  \label{recurH2}
  H_N^{(m|n)}(x,y)
  =
    \sum_{k=1}^N
  (-)^{k+1}
  \frac{
    (q ; q)_{N-1}
    }{
    (q ; q)_{N-k}
    } \cdot
  E_k(x,y) \cdot
  H_{N-k}^{(m|n)}(x,y)  .
\end{equation}
Notice that,
since
the elementary polynomials $E_k(x,y)$ do not vanish for
arbitrary $k$,
the polynomials $H_N^{(m|n)}(x,y)$ now depend on
every lower degree  polynomials
$H_j^{(m|n)}(x,y)$ for $j<N$
contrary to the first type recursion relation~\eqref{FirstRecur}.
In the case of $n=0$, both~\eqref{FirstRecur} and~\eqref{recurH2}
reduce to the recursion relation for the Rogers--Szeg{\"o}
polynomial~\cite{Hikam94d,Hikam94e}.

\subsection{Supersymmetric Skew Schur Polynomial}

We define  another $q$-polynomials
$F_N^{(m|n)}(x, y)
=
F_N^{(m|n)}(x_1, \dots, x_m , y_1, \dots , y_n ; q)
$
following Ref.~\citen{KiriKuniNaka96a};
\begin{equation}
  \label{defineF}
  F_N^{(m|n)}(x,y)
  =
  \sum_{r=1}^N
  \sum_{
    \substack{
      m_1 + \dots + m_r =N
      \\
      m_r \geq 1
      }}
  q^{
    \displaystyle
    \frac{N(N+1)}{2}
    -
    \sum_{i=1}^r
    \left(
      m_1 + \dots + m_i
    \right)
    }
  \times
  S_{\langle m_1 , \dots, m_r \rangle}(x,y) ,
\end{equation}
where $S_{\langle m_1 , \dots , m_r \rangle}(x,y)$ is the
supersymmetric skew Schur polynomial for the skew Young diagram
$\langle m_1 , \dots , m_r \rangle$~\eqref{border_strip}.
Being motivated from the expression of the 
skew Schur polynomials~\eqref{skewSchur} in non-supersymmetric case,
we define  the above mentioned supersymmetric   Schur polynomials 
through the supersymmetric elementary
polynomials~\eqref{susyElement}
as
\begin{equation}
  \label{Det_express}
  S_{\langle m_1 , \dots, m_r \rangle}(x,y)
  =
  \begin{vmatrix}
    E_{m_r} & E_{m_r + m_{r-1}} & \dots & \dots & E_{m_r +
      \dots + m_1}
    \\
    1 & E_{m_{r-1}} & E_{m_{r-1} + m_{r-2}}& \dots & E_{m_{r-1} +
    \dots + m_1}
  \\
  0 & 1 & E_{m_{r-2}}& \dots & \vdots
  \\
  \vdots & \ddots & \ddots & \ddots& \vdots
  \\
  0 & \dots & 0 & 1 & E_{m_1}
  \end{vmatrix} .
\end{equation}
Remark that, since $E_N(x,y) \neq 0$ for arbitrary $N$,
$S_{\langle m_1, \dots , m_r \rangle}(x,y)$ will be non-trivial for
arbitrary set of $m_1$, $m_2, \dots, m_r$.
Expanding the determinant in~\eqref{Det_express}
along the first row, we get a recursion relation for the
supersymmetric Schur polynomials as
\begin{equation*}
  S_{\langle m_1, \dots m_r \rangle}(x,y)
  =
  \sum_{i=1}^r
  (-)^{i+1} \,
  E_{m_r + \dots + m_{r-i+1}}(x,y) \cdot
  S_{\langle m_1, \dots, m_{r-i} \rangle}(x,y) .
\end{equation*}
By substituting the above  relation to the
expression~\eqref{defineF} and
following  Appendix~A in Ref~\citen{KiriKuniNaka96a},
we find
that the polynomial
$F_N^{(m|n)}(x,y)$ satisfies the recursion relation,
\begin{equation}
  F_N^{(m|n)}(x,y)
  =
  \sum_{k=1}^N
  (-)^{k+1}
  \frac{
    (q ; q)_{N-1}
    }{
    (q ; q)_{N-k}
    } \cdot
  E_k(x,y) \cdot
  F_{N-k}^{(m|n)}(x,y)   ,
\end{equation}
which is exactly same with the recurrence relation for the polynomials 
$H_N^{(m|n)}(x,y)$~\eqref{recurH2}.
We can easily recognize that the initial conditions are  same for two
sets of polynomials, $H_N^{(m|n)}(x,y)$ and $F_N^{(m|n)}(x,y)$,
\emph{e.g.},
\begin{align}
  F_0^{(m|n)}(x,y)
  & = 1 ,
  \nonumber
  \\[2mm]
  F_1^{(m|n)}(x,y)
  & = S_{[1]}(x,y) ,
  \nonumber
  \\[2mm]
  F_2^{(m|n)}(x,y)
  & =
  q \,  S_{[1^2]}(x,y) + S_{[2]}(x,y) ,
  \label{poly_example}
  \\[2mm]
  H_0^{(m|n)}(x,y)
  & = 1 ,
  \nonumber
  \\[2mm]
  H_1^{(m|n)}(x,y)
  & =
  \sum_i x_i + \sum_j y_j ,
  \nonumber
  \\[2mm]
  H_2^{(m|n)}(x,y)
  & =
  \sum_i x_i^{~2} + q \, \sum_j y_j^{~2}
  + (1+q)
  \Bigl(
  \sum_{i_1 < i_2} x_{i_1} \, x_{i_2}
  +  \sum_{i,j} x_i \, y_j
  + \sum_{j_1 < j_2} y_{j_1} \, y_{j_2}
  \Bigr) .
  \nonumber
\end{align}
As a result, we conclude  that
\begin{equation}
  \label{FequalH}
  F_N^{(m|n)} (x,y)
  = H_N^{(m|n)}(x,y) ,
\end{equation}
and that the polynomial $F_N^{(m|n)}(x=1,y=1)$  also gives  the
partition function $\mathcal{Z}^{(m|n)}(q)$ for the SP model.

\section{Motif: Eigenstates of SP Model}
\label{sec:representation}

We study eigenstates for the supersymmetric
su($m|n$) 
Polychronakos  model~\eqref{Hamilton_P}.
Due to the Yangian symmetry of the system,
we can use the   motif~\cite{HHTBP92,Hal94} which spans the Fock
space of the Yangian invariant spin systems.

The motif $d$ for ($N+1$)-site spin chain  is given by an
$N$ sequence of
0's and 1's;
$
d= (d_1, d_2, \dots , d_{N})
$
with
$d_j \in \{ 0, 1 \}$.
A value $d_j=1$
(resp. $d_j=0$)
denotes that the $j$-th energy level is occupied
(resp. empty)  by the quasi-particle.
The energy of the SP model  is given by
\begin{equation}
  \label{energy_motif}
  E(d) = \sum_{j=1}^{N} j \, d_j .
\end{equation}
We remark that the energy of the Yangian invariant Haldane--Shastry
type spin chain~\cite{Hal88,Sha88,HHTBP92} is given by
$E_{\text{HS}}
=\sum_j  j \, ( N-j) \, d_j
$.
We define $g_N(d)$ as a degeneracy of the motif $d=(d_1, \dots,d_N)$
for ($N+1$)-site spin chain.
We note that $g_N(d)$ generally depends on $x$ and $y$, and that
the degeneracy for each motif is given by setting
$x_i=y_j=1$.
Then the partition function~\eqref{partition} is written as
\begin{equation}
  \label{partition_motif}
  \mathcal{Z}_{N+1}(q)
  =
  \sum_d
  q^{\displaystyle
    E(d)
    }
  \,
  g_N(d) .
\end{equation}
In the su($m$) case, we have a selection rule for motifs
such that $m$-consecutive 1's are forbidden.
On the contrary, in the su($m|n$)
supersymmetric case we do not have such selection
rules;
every sequence $d$ is permitted generally~\cite{Hal94,BasuMa99a}.
{}From the first few polynomials~\eqref{poly_example} we can read
as
\begin{align*}
  g_0(~)
  & = S_{[1]}(x,y) =  e_1(x) + e_1(y) ,
  \\
  g_1(0)
  & =
  S_{[2]}(x,y) ,
  \\
  g_1(1)
  & =
  S_{[1^2]}(x,y) .
\end{align*}
Below we give two methods to obtain the degeneracy for the  general
motifs.

\subsection{Recursion Relation for Motif}

We study the representation of motif based on the first recursion
relation~\eqref{FirstRecur}.
Using~\eqref{partition_motif}, we can translate the recursion
relation~\eqref{FirstRecur} for the SRS polynomials to that for motifs.

We  give the several examples as follows.
By putting some special $m$ and $n$ to
the first recurrence relation~\eqref{FirstRecur} and translating it in 
terms of motif,
we get
the following equations;

\begin{enumerate}

  \def\labelenumi{(\Alph{enumi}).}

\item    su($2$);
  \begin{align}
    g_N(d_1 , \dots, d_N)
    & =
    \delta_{d_N, 0} \cdot e_1(x)
    \cdot g_{N-1}(d_1, \dots, d_{N-1})
    \nonumber \\
    & \qquad
    -
    \left(
      \delta_{d_N , 0} \cdot  \delta_{d_{N-1} , 0}
      -
      \delta_{d_N , 1} \cdot \delta_{d_{N-1} , 0}
    \right) \cdot e_2(x) \cdot
    g_{N-2}(d_1,\dots,d_{N-2}) .
  \end{align}

\item su($2|1$);
\begin{align}
  g_N(d_1, \dots, d_N)
  & =
  \left(
    \delta_{d_N,0} \, e_1(x) + \delta_{d_N , 1} e_1(y)
  \right) \cdot g_{N-1}(d_1, \dots, d_{N-1})
  \nonumber \\
  & \qquad
  -
  \left(
    \delta_{d_N , 0} \delta_{d_{N-1} , 0}
    -
    \delta_{d_N , 1} \delta_{d_{N-1} , 0}
  \right) \, e_2(x) \cdot
  g_{N-2}(d_1,\dots, d_{N-2}  ) .
\end{align}

\item su($2|2$);
  \begin{align}
    g_N (d_1, \dots, d_N)
    & =
    \left(
      \delta_{d_N,0} \, e_1(x) + \delta_{d_N, 1} \, e_1(y)
    \right) \cdot
    g_{N-1}(d_1, \dots, d_{N-1})
    \nonumber \\
    & \qquad
    -
    \Bigl(
      \bigl(
        \delta_{d_N , 0} \, \delta_{d_{N-1},0}
        -
        \delta_{d_N,1} \, \delta_{d_{N-1} , 0}
      \bigr) \, e_2(x)
      \nonumber \\
      & \qquad \qquad
      -
      \bigl(
        \delta_{d_N , 0} \, \delta_{d_{N-1},1}
        -
        \delta_{d_N,1} \, \delta_{d_{N-1} , 1}
      \bigr) \, e_2(y)
    \Bigr) \cdot
    g_{N-2}(d_1, \dots, d_{N-2}).
  \end{align}
\end{enumerate}
See that
the recurrence equation for 
the su($2|2$) case reduces to those for  su($2|1$) and su($2$) cases
by setting $e_2(y)=0$ and $e_1(y)=e_2(y)=0$, respectively.

By using these recurrence equations for the motifs, we can compute
degeneracy for each motif by setting $x_i = y_j = 1$.
We give some examples for $N=4$ below.
\begin{enumerate}
  \def\labelenumi{(\Alph{enumi}).}

\item su($2$) and su($1|1$);

  \bigskip

  \begin{center}
    \begin{tabular}{cccc}
      motif & energy & degeneracy for  su(2) & degeneracy for  su($1|1$) \\
      \hline \hline
      (000) & 0 & 5  & 2\\
      (100) & 1 & 3 & 2\\
      (010) & 2 & 4 & 2\\
      (001) & 3 & 3 & 2\\
      (101) & 4 & 1 & 2\\
      (110) & 3 & $-$ & 2 \\
      (011) & 5 & $-$ & 2 \\
      (111) & 6 & $-$ & 2 \\
      \hline \hline
      & \text{total} & $2^4$ & $2^4$
    \end{tabular}
  \end{center}

\item su(3) and su($2|1$);

  \bigskip

  \begin{center}
    \begin{tabular}{cccc}
      motif & energy & degeneracy for su($3$) & degeneracy for su($2|1$)  \\
      \hline \hline
      (000) & 0 & 15 &9 \\
      (100) & 1 & 15& 12 \\
      (010) & 2 & 21 & 16 \\
      (001) & 3 & 15& 12 \\
      (101) & 4 & 9 & 12 \\
      (110) & 3 & 3 & 8 \\
      (011) & 5 & 3 & 8 \\
      (111) & 6 & $-$ & 4 \\
      \hline \hline
      & \text{total} & $3^4$ & $3^4$
    \end{tabular}
  \end{center}

\end{enumerate}

\subsection{Motif and Skew Young Diagram}

We have shown that the representation of motif is given recursively by 
use of the first recursion relation~\eqref{FirstRecur}.
In this subsection, we shall show that,
by use of the skew Young
diagram and the   polynomials $F_N^{(m|n)}(x,y)$, we can directly give
the representation for motif.
We can then     give the decomposition rule for each motif, and clarify 
the hyper-multiplet structure in the spectrum of the SP model.

{}From a correspondence between a power of $q$ in~\eqref{defineF} and
the energy for motif $d$ in~\eqref{partition_motif}, we can define a
map from motif
$d$
to the skew Young diagram
$\langle m_1 ,\dots, m_r \rangle$~\cite{KiriKuniNaka96a}.
A rule for translation is:
we read a motif $d=(d_1, d_2, \dots)$
from the left, and
we add a box under (resp. left) the box when we encounter `$d_j=1$'
(resp. `$d_j=0$').
One sees that there is a one-to-one correspondence between motifs and
skew Young diagrams.
As we have  stressed before, the supersymmetric   skew Schur
polynomials
$S_{\langle m_1, \dots, m_r \rangle}(x,y)$ defined in~\eqref{Det_express} do
not vanish for arbitrary set of $m_1, \dots, m_r$,
and
this fact proves that
there is no selection rule 
for motif $d$ while the $m$-consecutive 1's are forbidden in the
non-supersymmetric
su($m$) case.

We give examples up to $N=4$ below.
We have  decomposed following a rule of the supersymmetric Young
diagrams.

\bigskip
\begin{center}
\begin{tabular}{cll}
  motif & skew Young diagram & decomposition \\
  \hline \hline
  (~) & $ \langle 1 \rangle$ & $[1]$ \\
  && \\
  (0) & $\langle 1,1 \rangle$ & $[2]$ \\
  (1) & $\langle 2 \rangle$ & $[1^2]$ \\
  && \\
    (11) & $\langle 3 \rangle$ & $[1^3]$ \\
    (01) & $\langle 1,2  \rangle$ & $[2,1]$ \\
    (10) & $\langle 2,1 \rangle$ & $[2,1]$ \\
    (00) & $\langle 1,1,1 \rangle$ & $[3]$  \\
   && \\
  (111) & $\langle 4 \rangle$ & $[1^4]$  \\
  (110) & $ \langle 3 , 1 \rangle$ & $[2,1^2]$ \\
  (101) & $ \langle 2 , 2 \rangle$ & $[2^2] \oplus [2, 1^2]$ \\
  (011) & $ \langle 1 , 3 \rangle$ & $[2,1^2]$ \\
  (100) & $ \langle 2 , 1 ,1 \rangle$ & $[3,1]$ \\
  (010) & $ \langle 1 , 2 , 1 \rangle$ & $[3,1] \oplus [2^2]$ \\
  (001) & $ \langle 1 , 1, 2 \rangle$ & $[3,1]$ \\
  (000) & $ \langle 1, 1, 1, 1 \rangle$ & $[4]$
\end{tabular}
\end{center}

\section{Distribution Function}
\label{sec:distribute}

In the preceding sections, we have shown  that the motifs, which are
the eigenstates of the SP model,
satisfy  some recursion relation.
As the motifs  are composed of the ``quasi-particles'', we
shall consider their  distribution function.
Once the recursion relation for the partition function is given, it is 
straightforward to compute the distribution function by use of the
asymptotic form of the partition function.
This method was originally introduced in
Refs.~\citen{Hikam95b,Hikam97e}
for a study of the exclusion
statistics~\cite{Hal91b}.
Namely in the SP model  the energy dispersion relation is
linear, so  we can  regard $q^k$ as
\begin{equation}
  V_k = \exp \left(  - \beta ( \varepsilon_k - \mu) \right)  ,
\end{equation}
in which $\beta$,  $\mu$, and $\varepsilon_k$ denote the inverse of
temperature, the chemical potential, and the $k$-th energy
respectively.
Correspondingly
the polynomial $H_N^{(m|n)}(x,y)$
can be viewed  as the \emph{restricted} partition function $\varphi_N$, 
in which particles can occupy up to $N$-th energy level
$\varepsilon_N$
(here   dispersion relation for energy $\varepsilon_k$ is arbitrary).
By assuming an asymptotic form of the restricted partition
function
\begin{equation}
  \label{Jost}
  \varphi_k \sim w^{-k},
\end{equation}
we find that the occupation of state $\varepsilon_k$ is given by
\begin{equation}
  \label{distribution}
  \langle n_{\text{av}} \rangle
  \simeq
  \frac{1}{k} \cdot
  \frac{\partial}{\partial ( \beta \, \mu)} \log \varphi_k
  =
  -
  \frac{\partial}{\partial ( \beta \, \mu)} \log w ,
\end{equation}
where the spectral parameter $w = w(V)$
is a function of $V=\exp \left( - \beta \, (\varepsilon - \mu) \right)$,
and it  can be determined from the recursion
relation for the restricted partition function.
In the same way,  we can compute the specific heat from the partition
function, and we get the central charge $c$ of the theory
as~\cite{FrenkSzene93a,Schou97a,BouwkSchou98a,BerkoBMcCo98a}
\begin{equation}
  \label{central_charge}
  c
  =
  - \frac{6}{\pi^2}
  \int_0^1 \mathrm{d}V \, \frac{1}{V} \log w(V)  ,
\end{equation}
which is often rewritten by use of the    Rogers' dilogarithm function
(see, \emph{e.g.}, Ref.~\citen{Kiril} and references therein).
In this section we demonstrate for some simple supersymmetric cases.

\begin{enumerate}
  \def\labelenumi{(\Alph{enumi}).}

\item
  su($1|1$);

  The recursion relation for the su($1|1$) SRS polynomials
  is read off as
  \begin{equation*}
    H_N^{(1|1)}(x,y)
    =
    \left(
      x + q^{N-1} \, y
    \right) \, H_{N-1}^{(1|1)}(x,y) , 
  \end{equation*}
  which shows that the restricted partition function satisfies
  \begin{equation}
    \varphi_N
    =
    \left( 1 + V_{N-1} \right) \,  \varphi_{N-1} .
  \end{equation}
  We see  using~\eqref{Jost} that the spectral parameter is given by
  \begin{equation}
    w = ( 1+ V)^{-1} .
  \end{equation}
  We thus obtain the central charge~\eqref{central_charge} as
  \begin{equation*}
    c=\frac{1}{2} ,
  \end{equation*}
  which denotes that the theory coincides with the free fermion
  theory.
  This fact is simply realized by explicitly computing the character
  in  \S~\ref{sec:WZW}.



\item
  su($m$);

  Based on a recursion relation for the RS polynomial, we see that the 
  restricted partition function $\varphi_k$ satisfies
  \begin{equation}
    \varphi_N
    =
    \sum_{i=1}^m (-)^{i-1} \,
    {}_m \mathrm{C}_i \, ( 1- V_{N-1}) \, (1 - V_{N-2}) \,
    ( 1-   V_{N-i+1}) \, \varphi_{N-i}  ,
  \end{equation}
  where ${}_m \mathrm{C}_i$ denote binomial coefficients.
  By assuming $V_N$ does not change drastically for $N$,
  we find~\cite{Hikam97e} that the spectral parameter $w$~\eqref{Jost}
  satisfies
  the functional equation,
  \begin{equation}
    V =
    \left(
      1 - ( 1 - V) \, w
    \right)^m .
  \end{equation}
  We also see
  that the central
  charge~\eqref{central_charge} is  computed
  as
  \begin{equation*}
    c = m-1 ,
  \end{equation*}
  which is known as the central charge for the level-1 su($m$) WZNW
  theory.

\item
  su($2|1$);

  The recursion relation for the partition function for the su($2|1$)
  SP model  is given as
  \begin{equation*}
    H_N^{(2|1)}(x,y) =
    \left( x_1 + x_2  + q^{N-1} \, y  \right) \cdot
    H_{N-1}^{(2|1)}(x,y)
    -
    (1 - q^{N-1}) \cdot x_1 \, x_2 \cdot
    H_{N-2}^{(2|1)}(x,y) ,
  \end{equation*}
  which, in terms of the restricted partition function, is rewritten as
  \begin{equation}
    \varphi_N
    = ( 2 + V_{N-1} ) \cdot \varphi_{N-1} - ( 1 - V_{N-1}) \cdot
    \varphi_{N-2} .
  \end{equation}
  By using an asymptotic form~\eqref{Jost}, we see that the spectral
  parameter satisfies the functional equation,
  \begin{equation*}
    (1-V) \, w^2 - (2+V) \, w + 1 = 0 ,
  \end{equation*}
  which gives
  \begin{equation}
    \langle n_{\text{av}}^{\mathrm{su}(2|1)} \rangle
    \simeq
    -
    \frac{\partial}{\partial ( \beta \, \mu)}
    \log
    \left(
      \frac{1}{2 \, (1 - V)}
      \left(
        2 + V - \sqrt{ V^2 + 8 \, V}
      \right)
    \right).
  \end{equation}
  We   obtain the central charge from~\eqref{central_charge}   as
  \begin{equation*}
    c = \frac{3}{2}= 1+ \frac{1}{2} ,
  \end{equation*}
  which  shows that the theory has one boson and one  fermion.

\item
  su($2|2$);

  We have the recursion relation for the restricted partition function as
  \begin{equation}
    \varphi_N
    =
    2 \, ( 1 + V_{N-1} ) \cdot \varphi_{N-1}
    - ( 1 - V_{N-1} ) \, ( 1 - V_{N-2}) \cdot \varphi_{N-2}  ,
  \end{equation}
  which gives the functional equation for the spectral
  parameter~\eqref{Jost} as
  \begin{equation*}
    (1-V)^2 \, w^2 - 2 \, (1+V) \, w + 1 = 0 ,
  \end{equation*}
  The average number of particle and the central charge are respectively 
  given as
  \begin{gather}
    \langle n_{\text{av}}^{\mathrm{su}(2|2)} \rangle 
    \simeq
    2 \,
    \frac{\partial}{\partial ( \beta \, \mu)}
    \log \left( 1 + \sqrt{V} \right)
    =
    2 \, \langle n^{\mathrm{su}(2)} \rangle
    ,
    \\[2mm]
    c  =  2  = 1+ 2 \times \frac{1}{2} . \nonumber
  \end{gather}
  This result is consistent with the fact that the theory includes one 
  boson and two fermions.

\item su($1|2$);

  The recursion relation for the restricted partition function is
  given by
  \begin{equation}
    \varphi_N
    =
    ( 1+ 2 \, V_{N-1} ) \, \varphi_{N-1}
    +
    ( 1 - V_{N-1} ) \, V_{N-2} \, \varphi_{N-2} .
  \end{equation}
  Correspondingly we get the functional equation for the spectral
  parameter~\eqref{Jost},
  \begin{equation*}
    V \, (V-1) \, w^2 - ( 1 + 2 \, V) \, w + 1 = 0 .
  \end{equation*}
  One sees a duality:
  $w^{\mathrm{su}(1|2)}(V) = V^{-1} \cdot w^{\mathrm{su}(2|1)}(V^{-1})$.
  The  distribution function~\eqref{distribution}
  and the central charge~\eqref{central_charge}
  are respectively computed as
  follows;
  \begin{gather}
    \langle n_{\text{av}}^{\mathrm{su}(1|2)} \rangle
    \simeq
    -
    \frac{\partial}{\partial ( \beta \, \mu)}
    \log
    \left(
      \frac{
        1 + 2 \, V - \sqrt{1 + 8 \, V}
        }{
        2 \, V \, (V-1)
        }
    \right) ,
    \\[2mm]
    c = 1 = 2 \times \frac{1}{2} ,
    \nonumber
  \end{gather}
  which shows that
  the su($1|2$)  theory has two independent fermions.
  We also find a duality for the occupation  of states as
  \begin{equation}
    \langle n_{\text{av}}^{\mathrm{su}(1|2)}(V) \rangle
    +
    \langle n_{\text{av}}^{\mathrm{su}(2|1)}(V^{-1}) \rangle
    =1 .
  \end{equation}
\end{enumerate}

\begin{figure}[htbp]
  \begin{center}
    \begin{psfrags}
      \psfrag{MMM}{$\varepsilon - \mu$}
      \psfrag{NNN}{$\langle n_{\text{av}} \rangle$}
      \psfig{file=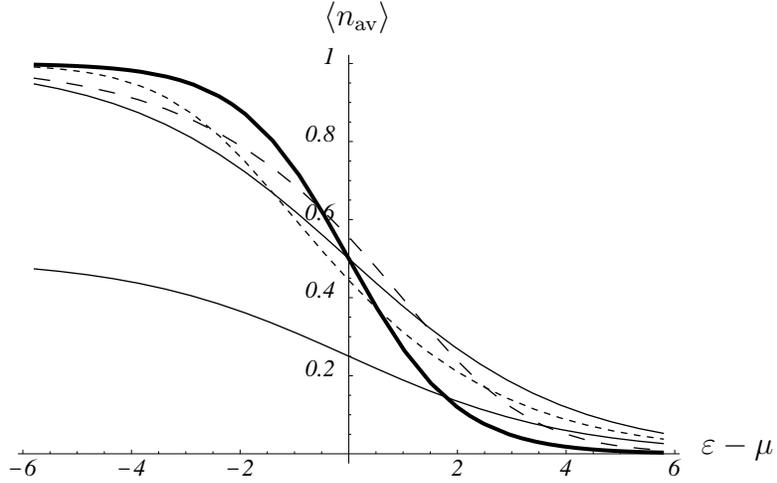}
    \end{psfrags}

    \caption{Average number of particle is plotted for free fermion
      and su($m|n$) cases ($m,n \leq 2$).
      A bold line is the distribution function for the free fermion,
      $\displaystyle \langle n_{\text{av}} \rangle = \frac{V}{1+V}$,
      and solid lines denote those for su(2) and su($2|2$).
      A dotted and a dashed line respectively show the distribution
      function
      for su($2|1$) and su($1|2$) cases.
      The energy is given in units $\beta^{-1}$.}
    \label{fig:su2}
  \end{center}
\end{figure}

In Fig.~\ref{fig:su2}
we plot an average occupation of states at a temperature $T>0$
for su($m|n$) cases ($m\leq 2$, $n=0,1,2$)
and for the free fermion case.
We see that
\begin{align*}
  \left.
    \langle n_{\text{av}} \rangle
  \right|_{ \varepsilon \to - \infty}
  & =
  \begin{cases}
    \frac{1}{2} ,
    & \text{for su($2$)},
    \\[2mm]
    1 ,
    & \text{for others},
  \end{cases}
  &
  \left.
    \langle n_{\text{av}} \rangle
  \right|_{\varepsilon \to \infty}
  & \simeq
  \begin{cases}
    \sqrt{V},  & \text{for su($2|2$)},
    \\[2mm]
    \frac{1}{\sqrt{2}}\sqrt{V},
    & \text{for su($2|1$)},
    \\[2mm]
    3 \, V ,
    &
    \text{for su($1|2$)},
  \end{cases}
\end{align*}
and that
\begin{align*}
  \left.
    \langle n_{\text{av}} \rangle
  \right|_{\varepsilon = \mu}
  & =
  \begin{cases}
    \frac{1}{2} ,  & \text{for su($2|2$)},
    \\[2mm]
    \frac{4}{9} ,
    & \text{for su($2|1$)},
    \\[2mm]
    \frac{5}{9},
    &
    \text{for su($1|2$)}.
  \end{cases}
\end{align*}

To close this section, we   observe that the central
charge~\eqref{central_charge}  for the su($m|n$) SRS polynomials
($m\neq 1$) will  be given generally as
\begin{equation}
  \label{center_general}
  c = m - 1 + \frac{n}{2} .
\end{equation}

\section{Character Formula}
\label{sec:WZW}

The SP model  has the supersymmetric Yangian
symmetry, and it is known that the first conserved operator  for the
Yangian operator is the Virasoro generator $L_0$.
We can thus conclude that the character
$\ch^{(m|n)}_{\Lambda}(q)$ for the
su$(m|n)_1$ WZNW model with the highest weight $\Lambda$
is related to the   SRS
polynomial as
\begin{equation}
  \label{chara_form}
  \ch^{(m|n)}_\Lambda(q)
  =
  \lim_{
    \substack{N\to \infty
      \\
      \text{condition}}
    }
  H_N^{(m|n)}(x=1, y=1 ; q ) .
\end{equation}

We do not know how to treat the formula~\eqref{chara_form} for general
cases, and
below we give some examples.

\begin{enumerate}
  \def\labelenumi{(\Alph{enumi}).}

\item 
su($1|1$);

It is easy to realize that the su($1|1$)
SRS polynomial is given
as the mere  $q$-product,
\begin{equation}
  H_N^{(1|1)}(x,y ; q )
  =
  \prod_{i=1}^N
  \left( x + q^{i-1} \, y \right)  .
\end{equation}
We then obtain the character formula
as
\begin{align}
  \ch^{(1|1)}_\Lambda (q)
  & =
  \lim_{N\to \infty}
  H_N^{(1|1)}(x=1,y=1;q)
  \nonumber \\
  & =
  \prod_{i=0}^\infty
  \left( 1 + q^i \right)   ,
\end{align}
which denotes that the su$(1|1)$  theory
coincides with  the free spinless   fermion theory as was demonstrated in
Ref.~\citen{Hal94}.

\item su($m$)~\cite{Hikam94d};

We have
\begin{equation}
  \ch_{\Lambda_j}^{(m)} (q)
  =
  \lim_{
    \substack{N \to \infty \\
      N \equiv j \mod m}
    }
  q^{\displaystyle
    \frac{m-1}{2 \, m}
    N^2
    }
  \cdot
  H_N^{(m|0)}(x=1, y=1; q^{-1}) ,
\end{equation}
which coincides with the character formula for su$(m)_1$ WZNW theory.



\item
su($1|n$);

Based on a numerical computation by use of
\textsc{Mathematica}
we suggest that
\begin{align}
  \ch^{(1|n)}_\Lambda (q)
  & =
  \lim_{N\to \infty}
  H_N^{(1|n)}(x_i=1,y=1;q)
  \nonumber \\
  & =
  \prod_{i=0}^\infty
  \left(
    1 + q^i
  \right)^n .
\end{align}
Indeed we can prove this identity  from the definition~\eqref{polyH} as
follows~\cite{Warnaar};
\begin{align*}
  \ch^{(1|n)}_\Lambda (q)
  & =
  \lim_{N \to \infty}
  \sum_{b_1, \dots, b_n \geq 0}
  \frac{(q;q)_N}{
    (q;q)_{b_1} \dots
    (q;q)_{b_n} \,
    (q;q)_{N-b_1- \dots - b_n}
    } \,
  q^{\displaystyle \frac{1}{2} \sum_{j=1}^n b_j \, (b_j - 1)}
  \\
  & =
  \sum_{b_1, \dots, b_n \geq 0}
  \frac{1}{
    (q;q)_{b_1} \dots
    (q;q)_{b_n}
    } \,
  q^{\displaystyle \frac{1}{2} \sum_{j=1}^n b_j \, (b_j - 1)}
  \\
  & =
  \left(
    \sum_{b=0}^\infty
    \frac{1}{(q;q)_b} \,
    q^{\frac{1}{2} b \, (b-1)}
  \right)^n
  \\
  & =
  \prod_{i=0}^\infty ( 1 + q^i )^n .
\end{align*}
This identity also supports that the theory is written by $n$
fermions~\eqref{center_general}.

\end{enumerate}

\section{Concluding Remarks}

We have studied eigenstates of the
supersymmetric extension of the
Polychronakos spin chain, which is the Yangian invariant system.
As it is known that the Yangian symmetry is realized in the level-1
WZNW theory~\cite{BPS94,Schou94},
the partition function can be regarded as the restricted character
formula.
We have introduced   the supersymmetric Rogers--Szeg{\"o} polynomials,
and
have shown that these polynomials give the partition function for the
SP model.
The SRS polynomials satisfy  two types of recursion
relations,~\eqref{FirstRecur} and~\eqref{recurH2};
the first   one  is useful to find out the number of degenerate
multiplets in each motif, while the second one gives us the
decomposition rule or hyper-multiplet structure of each motif by the
skew Young diagrams.
The first recursion relation  has  another aspect.
The spectrum~\eqref{energy_motif}    shows that
the motif denote the quasi-particle excitation of the SP model.
In fact the recursion relation   denotes  the exclusion statistics
of the quasi-particles, and
we have  computed  the distribution function following
Ref.~\citen{Hikam97e}.
One possible application of our results is for the edge states of the
quantum Hall effect.
As is well known from studies on the quantum Hall effect, the theory
of the  edge
state of the Laughrin state  is related to the Calogero--Sutherland
model, and the quasi-particles  have exclusion statistics.
Our supersymmetric theories should  also have close connections with  the
edge states.
Especially the $c=3/2$ su($2|1$) theory gives  quasi-hole excitations 
for the Pfaffian wave function~\cite{MoRe91}, and the
$c=2$ su($2|2$) theory is for the Haldane--Rezayi
states~\cite{HaldReza89}.
We hope to discuss in detail in a future issue.

It has been found earlier that, the partition function 
of the vertex model, associated with the vector representation of 
$U_q(\Hat{s\ell}_m)$  quantized affine algebra, exactly coincides 
with that of the $su(m)$ Polychronakos model~\cite{KiriKuniNaka96a}.
So, as
a future study, it might be interesting to investigate the partition function 
of the vertex model associated with the vector representation of 
$U_q(\hat{s\ell}_{m|n})$  algebra and examine whether such partition 
function also coincides with the polynomial 
partition function~\eqref{polyH}
of $su(m|n)$ SP
model. Moreover, in analogy with the non-supersymmetric
case~\cite{KiriKuniNaka96a},
the spectral decomposition of the above mentioned vertex model 
should be intimately connected with the supersymmetric skew 
Young diagrams and associated Schur functions which have been 
studied by us.
Such intriguing relations between the 
vertex model with nearest-neighbor type interaction and Polychronakos  
spin chain with long-ranged interaction may enlighten our understanding 
of the common integrable structure behind these models.
Another 
interesting problem might be to study the supersymmetric version of 
 one-row Macdonald polynomial.
 As is well known,
the one-row Macdonald polynomial can be viewed as 
a deformation of RS polynomial.
Thus one can obtain a realization of 
one-row Macdonald polynomial by deforming the spinon representation 
associated with $su(m)$ RS polynomial~\cite{Hikam96c}.
In analogy with this 
non-supersymmetric case, it should be possible to find out a spinon 
representation for our supersymmetric RS polynomial~\eqref{polyH}.
Furthermore,
by deforming such spinon representation in an appropriate way, 
one may obtain some novel realization of supersymmetric one-row 
Macdonald polynomial.

\section*{Acknowledgement}
The authors would like to thank Y.~Komori for  communications at
the early stage of this work.
One of the authors (BBM) thanks to Japan Society for the Promotion of
Science for a fellowship (JSPS-P97047) which supported  this work.

\newpage

\begin{thebibliography}{10}

\bibitem{Hal88}
F.~D.~M. Haldane: Phys. Rev. Lett. \textbf{60}, 635 (1988).

\bibitem{Sha88}
B.~S. Shastry: Phys. Rev. Lett. \textbf{60}, 639 (1988).

\bibitem{HHTBP92}
F.~D.~M. Haldane, Z.~N.~C. Ha, J.~C. Talstra, D.~Bernard, and V.~Pasquier:
  Phys. Rev. Lett. \textbf{69}, 2021 (1992).

\bibitem{Poly93a}
A.~P. Polychronakos: Phys. Rev. Lett. \textbf{70}, 2329 (1993).

\bibitem{Frah93}
H.~Frahm: J. Phys. A: Math. Gen. \textbf{26}, L473 (1993).

\bibitem{Poly94a}
A.~P. Polychronakos: Nucl. Phys. B \textbf{419}, 553 (1994).

\bibitem{Hikam94d}
K.~Hikami: Nucl. Phys. B \textbf{441}, 530 (1995).

\bibitem{BPS94}
D.~Bernard, V.~Pasquier, and D.~Serban: Nucl. Phys. B \textbf{428}, 612 (1994).

\bibitem{Schou94}
K.~Schoutens: Phys. Lett. B \textbf{331}, 335 (1994).

\bibitem{Hal94}
  F.~D.~M. Haldane: In \emph{Correlation Effects in Low Dimensional Electron
    Systems},
  edited by A.~Okiji and N.~Kawakami (Springer, Berlin, 1994)
  pp.~3--20.

\bibitem{Hikam94e}
K.~Hikami: J. Phys. Soc. Jpn. \textbf{64}, 1047 (1995).

\bibitem{Andre76}
G.~E. Andrews: \emph{The Theory of Partitions} (Addison-Wesley, London, 1976).


\bibitem{BiruUjinWada99a}
  B.~Basu-Mallick, H.~Ujino, and M.~Wadati:
preprint
(hep-th/9904167), to appear in J. Phys. Soc. Jpn.

\bibitem{Macdo95}
I.~G. Macdonald: \emph{Symmetric Functions and Hall Polynomials} (Oxford Univ.
  Press, Oxford, 1995), 2nd ed.

\bibitem{WFult97Book}
  W.~Fulton: \emph{Young Tableaux}.
  (Cambridge Univ.  Press, Cambridge, 1997).

\bibitem{BalanIBars82a}
A.~B. Balantekin and I.~Bars: J. Math. Phys. \textbf{23}, 1239 (1982).

\bibitem{BarMorRue83a}
I.~Bars, B.~Morel, and H.~Ruegg: J. Math. Phys. \textbf{24}, 22253 (1983).

\bibitem{KiriKuniNaka96a}
A.~N. Kirillov, A.~Kuniba, and T.~Nakanishi: Commun. Math. Phys. \textbf{185},
  441 (1997).

\bibitem{BasuMa99a}
B.~Basu-Mallick: Nucl. Phys. B \textbf{540}, 679 (1999).

\bibitem{Hikam95b}
K.~Hikami: Phys. Lett. A \textbf{205}, 364 (1995).

\bibitem{Hikam97e}
K.~Hikami: Phys. Rev. Lett. \textbf{80}, 4374 (1998).

\bibitem{Hal91b}
F.~D.~M. Haldane: Phys. Rev. Lett. \textbf{67}, 937 (1991).

\bibitem{FrenkSzene93a}
E. Frenkel and A. Szenes: Duke Math. J., IMRN, \textbf{2}, 53 (1993).

\bibitem{Schou97a}
K.~Schoutens: Phys. Rev. Lett. \textbf{79}, 2608 (1997).

\bibitem{BouwkSchou98a}
P.~Bouwknegt and K.~Schoutens: Nucl. Phys. B \textbf{547}, 501 (1999).

\bibitem{BerkoBMcCo98a}
  A.~Berkovich and B.~M. McCoy:
  in \emph{Statistical Physics on the Eve of the 21st Century},
  edited by M. T. Batchelor and L. T. Wille
  (World Scientific,  Singapore, 1999),
  pp.~240--256.

\bibitem{Kiril}
  A. N. Kirillov:
  Prog. Theor. Phys. Suppl. \textbf{118}, 61 (1994).

\bibitem{Warnaar}
  S. Ole Warnaar: private communication.
  \\
  We thank Dr. S. Ole Warnaar for pointing out this identity.

\bibitem{MoRe91}
G.~Moore and N.~Read: Nucl. Phys. B \textbf{360}, 362 (1991).

\bibitem{HaldReza89}
F.~D.~M. Haldane and E.~H. Rezayi: Phys. Rev. Lett. \textbf{60}, 956 (1989).

\bibitem{Hikam96c}
K.~Hikami: J. Phys. A: Math. Gen. \textbf{30}, 2447 (1997).

\end{thebibliography}

\end{document}